\documentclass[prb,twocolumn,showpacs,preprintnumbers,amsmath,amssymb]{revtex4}% Physical Review B
\usepackage{graphicx}% Include figure files
\usepackage{dcolumn}% Align table columns on decimal point
\usepackage{amsbsy}

\begin{document}

\title{The physical origin of the Fresnel drag of light by a moving dielectric medium }

\author{A.~Drezet}
\email{aurelien.drezet@uni-graz.at} \affiliation{Institut f\"ur
Experimentalphysik, Karl Franzens Universit\"at Graz,
Universit\"atsplatz 5 A-8010 Graz, Austria}

%\date{\today}% It is always \today, today,
             %  but any date may be explicitly specified

\begin{abstract}We present a new derivation of the
Fresnel-Fizeau formula for the drag of light by a moving medium
using a simple perturbation approach. We focus particulary on the
physical origin of the phenomenon and we show that it is very
similar to the Doppler-Fizeau effect. We prove that this effect is,
in its essential part, independent of the theory of relativity. The
possibility of applications in other domains of physics is
considered.
\end{abstract}

\pacs{03.~30.~+p; 32.~80.~-t;11.~80.~-m;42.~25.~-p} \maketitle

% DON'T CHANGE THIS LINE
\section{Introduction}
It is usual to consider the famous experiment of Fizeau
(1851)\cite{Fizeau} on the drag of light by a uniformly moving
medium as one of the crucial experiments which, just as the
Michelson-Morley experiment, cannot be correctly understood without
profound modification of Newtonian space-time concepts (for a review
of Einstein's relativity as well as a discussion of several
experiments the reader is invited to consult
\cite{Jackson,Bladel,Smith}). The result of this experiment which
was predicted by Fresnel\cite{Fresnel}, in the context of elastic
theory, is indeed completely justified by well known arguments due
to von Laue (1907) \cite{Laue}. He deduced the Fresnel-Fizeau result
for the light velocity $v$ in a medium, corresponding to a
relativistic first order expansion of the Einstein velocity
transformation formula:
\begin{equation}
v=\frac{\frac{c}{n_{0}}+ v_{e}}{1+\frac{v_{e}}{cn_{0}}}\simeq
\frac{c}{n_{0}}+v_{e}\left(1-\frac{1}{n_{0}^{2}}\right)+
O\left(\frac{v^{2}}{c^{2}}\right).\label{un}
\end{equation}
Here, $n_{0}$ represents the optical index of refraction of the
dielectric medium in its proper frame, and we suppose that the
uniform medium motion with velocity $v_{e}$ is parallel to the path
of the light and oriented in the same direction of propagation. In
the context of electromagnetic theory \cite{Minkowsky,Pauli} all
derivations of this effect are finally based on the invariance
property of the wave operator $\partial_{\mu}\partial^{\mu}[...]$ in
a Lorentz transformation.  It is easy to write the wave equation
$[\partial_{\mathbf{x}'}^{2}-\left(n_{0}^{2}/c^{2}\right)\partial_{t'}^{2}]\psi=0$
in the co-moving frame R'$\left(\mathbf{x'},t'\right)$ of the medium
in covariant form\cite{Leonhardt}
$[\partial_{\mu}\partial^{\mu}+\left(n_{0}^{2}-1\right)\left(v^{\nu}\partial_{\nu}\right)^{2}]\psi=0$
which is valid in all inertial frames and which for a plane waves,
implies the result of Eq.~\ref{un}. In this calculation we obtain
the result $v=c/n_{0}+v_{e}$ if we use the Galilean transformation
which proves the insufficiency of Newtonian dynamics.\\ However, the
question of the physical meaning of this phenomenon is not
completely clear. This fact is in part due to the existence of a
derivation made by Lorentz (1895) \cite{Lorentz} based on the mixing
%correction next line
between the macroscopic Maxwell's equations and a microscopic
electronic oscillator model which is classical in the sense of the
Newtonian dynamics. In his derivation Lorentz did not use the
relativistic transformation between the two coordinate frames:
laboratory and moving medium. Consequently, the relativistic nature
of the reasoning does not appear explicitly. Following the point of
view of Einstein (1915) \cite{Einstein} the Lorentz demonstration
must contain an implicit hypothesis of relativistic nature, however,
this point has not been studied in the literature. Recent
developments in optics of moving media
\cite{Leonhardt,Wilkens,Artoni,Yu} allows us to consider this
question as an important one to understand the relation between optics,
relativity and newtonian dynamics. This constitutes the subject of
the present paper. Here, we want to analyze the physical origin of
the Fresnel-Fizeau effect. In particular we want to show that this
phenomenon is, in its major part, independent of relativistic
dynamics.\\ The paper is organized as follows. In section II we
present the generalized Lorentz ``microscopic-macroscopic''
derivation of the Fresnel formula and the principal defect of this
treatment. In section III we show how to derive the Fresnel result
in a perturbation approach based on the Lorentz oscillator model and
finally in IV we justify this effect independently from all physical
assumptions concerning the electronic structure of matter.

\section{The Lorentz electronic model and its generalization} In this
part, we are going to describe the essential contents of the Lorentz
model and of its relativistic extension. Let
$\boldsymbol{\xi}\left(\boldsymbol{x},t\right)$ be the displacement
of an electron from its equilibrium position at rest, written as an
explicit function of the atomic position $\boldsymbol{x}$ and of the
time $t$. In the continuum approximation we can write the equation
of motion for the oscillator as $
\partial_{t}^{2}\boldsymbol{\xi}\left(\boldsymbol{x},t\right) +
\omega_{0}^{2}\boldsymbol{\xi}\left(\boldsymbol{x},t\right)\simeq
-\frac{e}{m} {\mathbf E}_{0}e^{-i\left(\omega
t-\boldsymbol{k}\cdot\boldsymbol{x}\right)}$ where the supposed
harmonic electric incident field appears and where the assumption of
small velocity allows us to neglect the magnetic force term. In the
case of a non relativistic uniformly moving medium we have
\begin{equation}
\left(\partial_{t}+\boldsymbol{v}_{e}\cdot\boldsymbol{\nabla}\right)^{2}\boldsymbol{\xi}
+ \omega_{0}^{2}\boldsymbol{\xi}\simeq -\frac{e}{m} \left({\mathbf
E}_{0}+\frac{\boldsymbol{v}_{e}}{c}\times\boldsymbol{B}_{0}\right)
e^{-i\left(\omega
t-\boldsymbol{k}\cdot\boldsymbol{x}\right)}\label{deux}\end{equation}
which includes the magnetic field
$\boldsymbol{B}=c\boldsymbol{k}\times\boldsymbol{E}/\omega$ of the
plane wave and the associated force due to the uniform motion with
velocity $\boldsymbol{v}_{e}$. The equation of propagation of the
electromagnetic wave in the moving medium has an elementary solution
when the velocity of the light and of the medium are parallel. If we
refer to a cartesian frame
$\boldsymbol{k}=k\hat{\boldsymbol{e}_{x}}$,
$\boldsymbol{v}_{e}=v_{e}\hat{\boldsymbol{e}_{x}}$ we have in this
case $\boldsymbol{E}=E_{0}e^{-i\left(\omega t-k x\right)}
\hat{\boldsymbol{e}}_{y}$, $ \boldsymbol{B}=c\frac{k}{\omega}
E_{0}e^{-i\left(\omega t-k x\right)} \hat{\boldsymbol{e}}_{z}$ for
the electromagnetic field and
\begin{eqnarray}
\boldsymbol{\xi}=-\frac{e}{m}\frac{\boldsymbol{E}_{0}\left(1-\frac{kv_{e}}{\omega
}\right)}{\omega_{0}^{2}-\left(\omega-k v_{e}\right)^{2}}
e^{-i\left(\omega t-k x\right)}\label{trois}
\end{eqnarray} for the displacement vector parallel to the $y$ axis.
 The relativistic extension of
 this model can be obtained directly
 putting $v_{e}=0$  in Eq.~\ref{deux} or \ref{trois} and  using a Lorentz
 transformation between the moving frame and the laboratory one. We deduce the displacement
\begin{eqnarray}
\boldsymbol{\xi}=-\frac{e}{m}\gamma_{e}\frac{\boldsymbol{E}_{0}\left(1-\frac{kv_{e}}{\omega
}\right)}{\omega_{0}^{2}-\gamma_{e}^{2}\left(\omega-k
v_{e}\right)^{2}} e^{-i\left(\omega t-k x\right)}\label{quatre}
\end{eqnarray} where
$\gamma_{e}=1/\sqrt{\left(1-v_{e}^{2}/c^{2}\right)}$. We could
alternatively obtain the same result considering the generalization
of the Newton dynamics i.~e.~ by doing the substitutions
$m\rightarrow m\gamma_{e}$ and
$\omega_{0}\rightarrow\omega_{0}\gamma_{e}^{-1}$ in Eq.~\ref{deux}.
The dispersion relation is then completely fixed by the Maxwell
equation $\frac{\partial^{2}}{\partial
x^{2}}E-\frac{1}{c^{2}}\frac{\partial^{2}}{\partial
t^{2}}E=\frac{4\pi}{c^{2}}\frac{\partial}{\partial t} J$, where the
current density $J$ is given by  the formula
$J=-eN\left(\partial_{t}+v_{e}\partial_{x}\right)\xi$ depending on
the local number of atoms per unit volume $N$ supposed to be
constant. Using $J$ and Eq.~\ref{trois} or \ref{quatre} we obtain a
dispersion relation $k^{2}=n^{2}\left(\omega\right)\omega^{2}/c^{2}$
%corrections next two lines
where the effective refractive index $n\left(\omega\right)$ depends
on the angular frequency $\omega$ and on the velocity $v_{e}$. The more general
index obtained using Eq.~\ref{quatre} is defined by the implicit
relation
\begin{equation}
n^{2}\left(\omega\right)=1+\gamma_{e}^{2}[n_{0}^{2}\left(\omega'\right)-1]
[1-\frac{n\left(\omega\right)v_{e}}{c}]^{2}\label{cinq}.
\end{equation}
Here $\omega'= \omega\left(1-\frac{n v_{e}}{c}\right)\gamma_{e}$,
and $n^{2}_{0}\left(\omega\right)=1+ 4\pi
N_{0}e^{2}/\left(\omega_{0}^{2}-\omega^{2}\right)/m $ is the
classical Lorentz index (also called Drude index) which contains the
local proper density
 which is defined in the frame where the medium is immobile by $N_{0}=N\gamma_{e}^{-1}$. These relativistic equations imply
directly the correct relativistic formula for the velocity of light
in the medium: Writing
$n_{0}^{2}-1=(n^{2}-1)(1-v_{e}^{2}/c^{2})/(1-nv_{e}/c)^{2}=
(n-v_{e}/c)^{2}/(1-nv_{e}/c)^{2}-1$ we deduce
\begin{equation}
\frac{c}{n_{0}}=\frac{c/n-v_{e}}{1-\frac{v_{e}}{cn}}.
\end{equation} which can be easily transformed into
\begin{equation}
v=\frac{c}{n}=\frac{c/n_{0}\left(\omega'\right)+v_{e}}{1+\frac{v_{e}}{cn_{0}\left(\omega'\right)}}.
\label{six}
\end{equation}
It can be added that by combining these expressions we deduce the
explicit formula
\begin{equation}
n^{2}\left(\omega\right)=1+\gamma_{e}^{-2}\frac{[n_{0}^{2}\left(\omega'\right)-1]}{
[1+\frac{n_{0}\left(\omega'\right)v_{e}}{c}]^{2}}.
\end{equation}
 The non relativistic case can be obtained directly from Eq.~\ref{trois} or by writting
 $\gamma_{e}=1$ in Eqs.~\ref{cinq},\ref{six}. This limit
\begin{equation}
v=\frac{c}{n}\simeq\frac{c}{n_{0}}+v_{e}[1-\frac{1}{n_{0}^{2}}+\omega\frac{d\ln
n_{0}}{d\omega}]+ O\left(\frac{v_{e}^{2}}{c^{2}}\right)
\end{equation}
is the Fresnel-Fizeau formula corrected by a
``frequency-dispersion" term due to Lorentz\cite{Lorentz}. For our
purpose, it is important to note that in the non-relativistic
limit of Eq.~\ref{cinq} we can always write the equality
\begin{equation}
\frac{c}{n}=\frac{c-v_{e}}{n'}+v_{e}
\end{equation}
where $n'=n\left(1-v_{e}/c\right)/\left(1-nv_{e}/c\right)$ is the
index of refraction defined relatively to the moving medium. We then
can see directly that the association of Maxwell's equation with
Newtonian dynamics implies a modification of the intuitive
assumption ``$c/n_{0}+v_{e}$" used in the old theory of emission. In
fact, the problem can be understood in the Newtonian mechanics using
the absolute time $t=t'$ and the transformation $x=x'+v_{e}t'$. In
the laboratory frame the speed of light, which in vacuum is $c$,
becomes $c/n_{0}$ in a medium  at rest. In the moving frame the
speed of light in vacuum is now $c-v_{e}$\cite{Jackson}. However,
due to invariance of acceleration and resultant force in a galilean
transformation we can interpret the presence of the magnetic term in
Eq.~\ref{deux} as a correction to the electric field in the moving
frame. This effective electric field affecting the oscillator in the
moving frame is then transformed into $E\left(1-nv_{e}/c\right)$. It
is this term which essentially implies the existence of the
effective optical index $n'\neq n_{0}$ and the light speed
$\left(c-v_{e}\right)/n'$ in the moving frame. It can be observed
that naturally Maxwell's equations are not invariant in a Galilean
transformation. The interpretation of $E\left(1-nv_{e}/c\right)$ as
an effective electric field is in the context of Newtonian dynamics
only formal: This field is introduced as an analogy with the case
$v_{e}=0$ only in order to show that $n'$ must be different from
$n_{0}$.
%%%%%%%%
%%%%%
%%%%%
%%%%%%%%
\section{Perturbation approach and optical theorem}
\begin{figure}
  \includegraphics[width=3in]{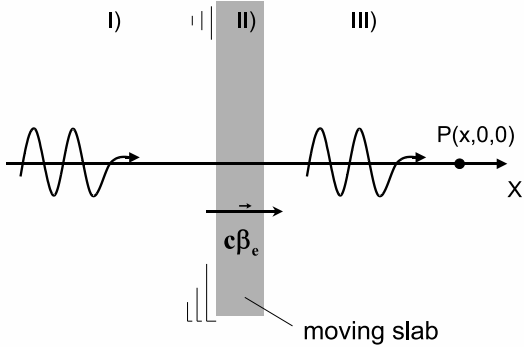}\\
  \caption{Representation of a linearly polarized electromagnetic
  plane wave travelling in a moving slab perpendicular to the $x$
  axis.  The velocity of the slab is $c\beta_{e}$, and the three
  spatial regions in front, in and after the slab are denoted by
  1, 2, and 3, respectively. We have plotted in addition a typical
  observation point P($x$,0,0). }\label{slab} \end{figure}

The difficulty of the preceding model is that the Lorentz derivation
does not clarify the meaning of the Fresnel-Fizeau phenomenon.
Indeed we justify Eq.~\ref{un} using a microscopical model which is
in perfect agreement with the principle of relativity. However we
observe that at the limit $v_{e}\ll c$ the use of the non
%corrections lignes +1, +2, +4 et +5
relativistic dynamics of Newton (see Eq.~\ref{deux}) gives the same
result. More precisely one can see from Eq.~\ref{deux} that the
introduction of the magnetic force
$-e\boldsymbol{v}_{e}\times\boldsymbol{B}/c$ in addition to the electric force
is already sufficient to account for the Fresnel-Fizeau effect and this even
if the classical force formula $\mathbf{F}=m\ddot{\mathbf{x}}(t)$ is
conserved. Since the electromagnetic force contains the ratio
$v_{e}/c$ and
 originates from Maxwell's equations this is already a term of relativistic
nature (Einstein used indeed this fact to modify the dynamical laws
of Newton\cite{Einstein2}). The derivation of Lorentz is then based
%corrections lignes +1,+5, +9, +10
on Newton as well as on Einstein dynamics. It is well know in
counterpart that the Doppler-Fizeau effect, which includes the same
factor $1-v_{e}/c$, can be understood without introducing Einstein's
relativity. Indeed this effect is just a consequence of the
invariance of the phase associated with a plane wave when we apply a
Galilean transformation (see \cite{Jackson}, Chap.~11) as well as a
Lorentz transformation. We must then analyze further in detail the
interaction of a plane wave with a moving dipole in
order to see if the Fresnel phenomenon can be understood independently of the specific Lorentz dynamics.\\
 We consider in
this part a different calculation based on a perturbation method and
inspired by a derivation of the optical theorem by
Feynman\cite{Feynman,Hulst}. Consider a thin slab of thickness $L$
perpendicular to the $x$ axis. Let this slab move along the positive
$x$ direction with the constant velocity
$v_{e}\boldsymbol{\hat{e}}_{x}$. Let in addition ${\mathbf
E}_{0}e^{-i\omega\left(t -x/c\right)}$ be the incident electric
field of a plane wave which pursues the moving slab (see
Fig.~\ref{slab}). Therefore, the electric field after the slab can
be formally written as
\begin{equation}
{\mathbf E}_{\text{after}}={\mathbf
E}_{0}e^{-i\omega\left(t-\delta t_{v_{e}}
-x/c\right)},\label{caput}
\end{equation}
where $\delta t_{v_{e}}$ appear as a retardation time produced by
the interaction of light with the slab and where all reflections are
neglected ($ \|{\mathbf E}_{\text{after}}\|=\|{\mathbf
E}_{\text{before}}\|$). For a ``motionless'' slab (i.~e., the case
considered by Feynman) we can write the travel time of the light
through the slab as $\Delta\tau_{0}=L_{0}/c+ \delta t_{0}=n_{0}\cdot
L_{0}/c$ and therefore $\delta t_{0}=\left(n_{0}-1\right)\cdot
L_{0}/c$ where $L_{0}$ defines the proper length of the slab in the
frame where it is at rest. For the general case of a moving slab of
reduced length $L=L_{0}\gamma_{e}^{-1}$ we find for the travel time:
\begin{equation}
\Delta\tau_{e}=\frac{\left(L+v_{e}\Delta\tau_{e}\right)}{c}+\delta
t_{v_{e}}=n\cdot\frac{\left(L+v_{e}\Delta\tau_{e}\right)}{c}\label{ultime}
\end{equation}
 and therefore the perturbation time is
 \begin{eqnarray}
\delta
t_{v_{e}}=\frac{\left(n-1\right)L}{\left(c-nv_{e}\right)}.\label{ultime2}
\end{eqnarray}
We can obtain this result more rigorously  by using Maxwell's
boundary conditions at the two moving interfaces separating the
matter of the slab and the air (see Appendix A).

 In order to evaluate the diffracted field which is  $\mathbf{E}_{\text{after}}-\mathbf{E}_{\text{before}}$ we can limit our calculation to a
 first order approximation. Thereby, each dipole of the
 Lorentz model as discussed above can be considered as being excited
 directly by the incident electromagnetic wave and where we can neglect all phenomena
  implying multiple interactions between light and matter. In this limit Eq.~\ref{caput} reduces to
\begin{eqnarray}
{\mathbf E}_{\text{after}}={\mathbf E}_{0}e^{-i\omega\left(t
-x/c\right)}e^{+i\omega \left(n-1\right)\case{L}{c-nv_{e}}} & &
\nonumber\\ \simeq{\mathbf E}_{0}e^{-i\omega\left(t -x/c\right)}
\left(1+i\omega\left(n-1\right)\frac{L}{c-nv_{e}}\right). & &
\label{trucmuch}
\end{eqnarray}
 If the distance between the slab and an observation point is much
 larger than $L$ we can consider the slab as a 2D continuous
 distribution of radiating point dipoles. The vector potential
 ${\mathbf A}_{\text{rad}}$ radiated by a relativistically moving point
 charge $e$ is in according with the Lienard-Wiechert's formula given
 by \cite{Jackson}:
\begin{eqnarray}
\mathbf{A}_{\text{rad}}\left(\mathbf{x},t\right)=
e\frac{\mathbf{v}/c}{\left(1-\mathbf{\hat{R}}\cdot \mathbf{\beta}\right)R} |_{ret}. \label{wiechert}
\end{eqnarray}
Here $R=\| \mathbf{ x}- \mathbf{ x}_{0}\left(t\right)\|$ is the distance separating the observation point $\mathbf{ x}$
(denoted by P) and the point charge position located at $\mathbf{ x}_{0}\left(t\right)$ at the time $t$; additionally
 $\boldsymbol{v}\left(t\right)=\dot{\boldsymbol{x}}_{0}\left(t\right)$ is the
velocity of the point charge and
$\boldsymbol{\hat{R}}\left(t\right)$ is
 the unit vector $\left(\mathbf{ x}- \mathbf{ x}_{0}\left(t\right)\right)/R\left(t\right)$. In this formula,
 in agreement with causality, all point charge variables are evaluated at the retarded time $t_{ret}=t-R\left(t_{ret}\right)/c$.\\
In the present case the motion of the point charge can be decomposed
into a uniform longitudinal component $\mathbf{ v}_{e}t$ oriented
along the positive $x$ direction and into a transversal oscillating
part $\mathbf{ \xi}\left(t\right)=\mathbf{
\xi}_{0}e^{-i\omega\left(1-\case{v_{e}}{c}\right)t}$ obeying the
%corrections lignes +1,+7,+8
condition $\|\dot{\mathbf{ \xi}}\left(t\right)\|/c\ll 1$. Owing to
this condition we can identify  $\left(1-\mathbf{
\hat{R}}\cdot\mathbf{ v}/c\right)$ with $\left(1-\mathbf{
\hat{R}}\cdot\mathbf{ v_{e}}/c\right)$. Consequently, in the
far-field the contribution of the electron uniform velocity is
cancelled by the similar but opposite contribution associated with
the nucleus of the atomic dipole: Only the vibrating contribution of
the electron survives at a long distance from the diffraction source.
If we add the contribution of each dipole of the slab acting on the
observation point P  at the time $t$ we obtain then the total
diffracted vector potential $\mathbf{ A}_{\text{diff}}$ produced by
the moving medium:
\begin{eqnarray}
\mathbf{ A}_{\text{diff}}\left(\mathbf{ x},t\right)\simeq -2\pi \gamma_{e}N_{0}L i\omega \frac{e}{c} \left(1-\beta_{e}\right)\mathbf{ \xi}_{0}\nonumber\\
\int_{0}^{+\infty}\rho d\rho\frac{
e^{-i\omega\left(1-\beta_{e}\right)\left(t-R\left(t_{ret}\right)/c\right)}}{\left(1-\boldsymbol{\hat{R}}\left(t_{ret}\right)\cdot\boldsymbol{v}_{e}/c\right)R\left(
t_{ret}\right)} ,\label{field}
\end{eqnarray}
Here $\rho$ is the radial coordinate in a cylindrical coordinate
system using the direction $x$ as a revolution axis, and the
quantity $\gamma_{e}N_{0}L 2\pi\rho d\rho$ is the number of dipoles
contained in the cylindrical volume of length $L$ and of radius
varying between $\rho$ and $\rho +d\rho$ if we consider a local
dipole density given by $\gamma_{e}N_{0}$. In this formula the
retarded distance $R\left(t_{ret}\right)$ is a function of $\rho$
%correction +1
and we have (see  the textbook of Jackson \cite{Jackson})
\begin{equation}
R\left(t_{ret}\right)=\gamma_{e}^{-1}\left(1-\mathbf{
\hat{R}}\left(t_{ret}\right)\cdot\mathbf{ v}_{e}/c
\right)^{-1}\sqrt{\rho^{2}+
\gamma_{e}^{2}\left(x-v_{e}t\right)^{2}}.\label{cacatoes}
\end{equation}
 This expression shows that the minimum $R_{min}$ is obtained for a
 point charge on the $x$ axis, and that:
\begin{equation}
 R_{min}=\left(x-v_{e}t\right)/\left(1-\beta_{e}\right).
\end{equation}
In order to evaluate the integral in Eq.~\ref{field} we must use
in addition the following relation (see Appendix C):
\begin{equation}
R\left(t_{ret}\right)=
\gamma_{e}^{2}\beta_{e}\left(x-v_{e}t\right)+
\gamma_{e}\sqrt{\rho^{2}+\gamma^{2}_{e}\left(x-v_{e}t\right)^{2}}\label{cactus}
\end{equation}
Hence, we obtain the following integral :
  \begin{eqnarray}
\mathbf{ A}_{\text{diff}}\left(\mathbf{ x},t\right)\simeq -2\pi i\omega\gamma_{e}^{2}N_{0}L\frac{e}{c} \left(1-\beta_{e}\right)\mathbf{ \xi}_{0}\nonumber\\\cdot e^{-i\omega\left(1-\beta_{e}\right)\left(t- \gamma_{e}^{2}\beta_{e}\left(x-v_{e}t\right)\right)}\nonumber\\
\cdot\int_{0}^{+\infty}\rho d\rho\frac{
e^{i\frac{\omega}{c}\left(1-\beta_{e}\right)\gamma_{e}\sqrt{\rho^{2}+\gamma^{2}_{e}\left(x-v_{e}t\right)^{2}}\}}}{
\sqrt{\rho^{2}+\gamma_{e}^{2}\left(x-v_{e}t\right)^{2}} }
,\label{fieldbis}
\end{eqnarray}
 where we have used the relations Eq.~\ref{cacatoes}, Eq.~\ref{cactus} in the denominator and in the
 exponential argument of the right hand side of  Eq.~\ref{field}, respectively. The diffracted field is
 therefore directly
 calculable by using the variable $u=
 \sqrt{\rho^{2}+\gamma_{e}^{2}\left(x-v_{e}t\right)^{2}} $. We obtain
 the result
\begin{eqnarray}
\mathbf{ A}_{\text{diff}}\simeq 2\pi \gamma_{e}L N_{0}
\frac{e}{c}\mathbf{ \xi}_{0}e^{-i\omega\left(t-x/c\right)}.
\end{eqnarray}
The total diffracted electric field $\mathbf{ E}_{\text{diff}}$ is
obtained using Maxwell's formula $\mathbf{ E}=-\left(1/c \right)
\partial_{t}\mathbf{ A}$, which gives:
  \begin{eqnarray}
\mathbf{ E}_{\text{diff}}\simeq 2\pi i\gamma_{e}L N_{0}
\omega\frac{e}{c}\mathbf{ \xi}_{0}e^{-i\omega\left(t-x/c\right)}.
\label{chose} \end{eqnarray}
The final result is given substituting
Eq.~\ref{quatre} in Eq.~\ref{chose} and implies by comparison with
Eq.~\ref{trucmuch}\begin{eqnarray}
 n\simeq 1+2\pi N_{0}\gamma_{e}^{2}\frac{e^{2}}{m}
\frac{ \left(1-\frac{ v_{e}}{c}\right)^{2}} {\omega_{0}^{2}-
\gamma_{e}^{2}\omega^{2}\left(1- \frac{v_{e}}{c}\right)^{2}}
.\label{ultime3}
\end{eqnarray}
%correction lignes +1, +3, +4, +6
This equation constitutes the explicit limit $N_{0}\rightarrow 0$
of Eq.~\ref{cinq} and implies the correct velocity formula
Eq.~\ref{six} when we neglect terms of
$\textrm{O}[N_{0}^{2}]$. It can again be observed that the present
calculation can be reproduced in the non relativistic case by
neglecting all terms of order $(v_{e}/c)^{2}$.

\section{Physical meaning and discussion}
The central fact in this reasoning is ``the travel condition" given
by Eqs.~\ref{ultime},\ref{ultime2}. Indeed, of the same order in
power of $N_{0}$ we can deduce the relation \begin{equation} \delta
t_{v_{e}}=\gamma_{e}\delta t_{0}\left(1-v_{e}/c\right)\end{equation}
and consequently the condition Eq.~\ref{ultime} reads
\begin{equation} \Delta\tau_{e}=\frac{L+v_{e}\left(\Delta\tau_{e}-\delta
t_{0}\gamma_{e}\right)}{c}+  \delta
t_{0}\gamma_{e}=n\cdot\frac{\left(L+v_{e}\Delta\tau_{e}\right)}{c}\label{ultime3}.
\end{equation}
If we call $\delta t_{0}$ the time during which the energy contained
in a plane of light moving in the positive x direction is absorbed
by the slab at rest in the laboratory, $\delta t_{0}\gamma_{e}$ is
evidently the enlarged time for the moving case. During the period
where this plane of light is absorbed by the slab its energy moves
at the velocity $v_{e}$. This fact can be directly deduced of the
energy and momentum conservation laws. Indeed, let $M
\gamma_{e}v_{e}$ be the momentum of the slab of mass $M$ before the
collision and $\epsilon$ the energy of the plane of light, then
during the interaction the slab is in a excited state and its energy
is now $E^{*}=\epsilon+M \gamma_{e}c^{2}$ and its momentum
$P^{*}=\epsilon/c+M \gamma_{e}v_{e}$. The velocity of the excited
slab is defined by $w=c^{2}P^{*}/E^{*}$ and we can see that in the
approximation $M\rightarrow \infty$ used here $w\simeq v_{e}$ (we
neglect the recoil of the slab). During $\delta t_{0}\gamma_{e}$ the
slab moves along a path length equal to $v_{e}\delta
t_{0}\gamma_{e}$ and thus the travel condition of the plane of
energy in the moving slab can be written
\begin{equation}
c \left(\Delta \tau_{e}- \gamma_{e}\delta
t_{0}\right)=L+v_{e}\left(\Delta \tau_{e}- \gamma_{e}\delta
t_{0}\right)\label{condition},
\end{equation}
which is an other form for Eq.~\ref{ultime3}. Now eliminating
directly $\Delta \tau_{e}$ in Eq.~\ref{condition} give us the
velocity $v$ of the wave:
\begin{equation}
 v=v_{e}+\frac{c-v_{e}}{1+\frac{c  \delta
t_{0}}{L}\left(1-\beta_{e}\right)\gamma_{e}},\label{fin}
 \end{equation}
 i.~e.~
 \begin{equation}
 v=v_{e}+\frac{c-v_{e}}{1+\left(n_{0}-1\right)\left(1-\beta_{e}\right)\gamma_{e}^{2}},
 \end{equation}
which depends on the optical index $n_{0}=1+c\delta t_{0}/L_{0}$.
After straightforward manipulations this formula becomes
\begin{equation}
v=\frac{c/n_{0}+v_{e}}{1+\frac{v_{e}}{cn_{0}}}\label{relativist}
\end{equation} which is the Einstein formula containing the Fresnel result as the limit behavior for small $v_{e}$.\\
It can be observed that this reasoning is even more natural if we
think in term of particles. A photon moving along the axis $x$ and
pursuing an atom moving at the velocity $v_{e}$ constitutes a good
analogy to understand the Fresnel phenomenon. This analogy is
evidently not limited to the special case of the plane wave
$e^{i\omega\left(t-x/c\right)}$. If for example we consider a small
wave packet which before the
   interaction with the slab has the form
\begin{equation}
E_{\text{before}}\left(x,t\right)=\int_{\Delta
\omega} d\omega a_{\omega}e^{i\left(kx-\omega t\right)},
\end{equation}
where $\Delta \omega$ is a small interval centered on
$\omega_{m}$, then after the interaction we must have:
\begin{equation}
E_{\text{after}}\left(x,t\right)=\int_{\Delta
\omega} d\omega a_{\omega}e^{i\left(kx-\omega [t-\delta
t\left(\omega\right)\right)},
\end{equation}
where $\delta t\left(\omega\right)$ is given by Eq.~\ref{ultime2}.
After some manipulation we can write these two wave packets in the
usual approximative form:
\begin{eqnarray}
E_{\text{before}}\simeq e^{i\left(k_{m}x-\omega_{m}t\right)}\int_{\Delta \omega} d\omega a_{\omega}e^{-i\left(\omega-\omega_{m}\right)[t-\partial k/\partial\omega_{m}x]} \nonumber\\
= e^{i\left(k_{m}x-\omega_{m}t\right)}
F\left(t-x/v_{g}\right)\nonumber
\end{eqnarray}
\begin{eqnarray}
E_{\text{after}}\simeq e^{i\left(k_{m}x-\omega_{m}[t-\delta
t\left(\omega_{m}\right)]\right)} F\left(t-x/v_{g}-\delta
t_{g}\right).
\end{eqnarray}
Here, $v_{g}=\partial\omega_{m}/\partial k_{m}=c$ is the group velocity of the pulse in vacuum and
 $ \delta t_{g}=\partial\left(\omega_{m}\delta t\left(\omega_{m}\right)\right)/\partial \omega_{m}$ is the
perturbation time associated with this group motion. This equation
for $F$ possesses the same form as Eq.~\ref{caput} and then the same
analogy which implies Eq.~\ref{ultime3} is possible. This can be
seen from the fact that we have
\begin{equation}
\delta t\left(\omega\right)=\gamma_{e}\delta
t_{0}(\omega')\left(1-v_{e}/c\right)\label{phase}
\end{equation} with $\omega'=\gamma_{e}\omega(1-v_{e}/c)$.
We deduce indeed
\begin{equation}
\delta t_{g}=\gamma_{e}\delta
t_{0g}(\omega'_{m})\left(1-v_{e}/c\right)\label{groupe},
\end{equation}
where we have $\delta
t_{0g}(\omega'_{m})=\partial\left(\omega_{m}\delta
t_{0}\left(\omega'_{m}\right)\right)/\partial \omega_{m}$
 i.~e.~$\delta t_{0g}(\omega'_{m})=\partial\left(\omega'_{m}\delta
t_{0}\left(\omega'_{m}\right)\right)/\partial \omega'_{m}$. Since
Eq.~\ref{phase} and Eq.~\ref{groupe} have the same form the Fresnel
law must be true for the group velocity.\\ It is important to remark
that all this reasoning conserves its validity if we put
$\gamma_{e}=1$ and if we think only in the context of Newtonian
%correction +1
dynamics. Since the reasoning with the travel time does not
explicitly use the structure of the medium involved (and no more the
magnetic force $-e\boldsymbol{v}_{e}\times\boldsymbol{B}/c$) it must
be very general and applicable
in other topics of physics concerning for example elasticity or sound. \\
Consider as an illustration the case of a cylindrical wave guide
with revolution axis $x$ and of constant length $L$ pursued by a
wave packet of sound. We suppose that the scalar wave $\psi$ obeys
the equation $[c^{2}\partial^{2}/\partial \mathbf{ r}^{2}-
\partial^{2}/\partial t^{2}]\psi=0$ where $c$ is the constant sound
velocity. The propagative modes in the cylinder considered at rest
in the laboratory are characterized by the classical dispersion
relation \begin{equation} \omega^{2}/c^{2}
=\gamma_{n,m}^{2}+k_{x}^{2}\label{dispersion}\end{equation} where
the cut off wave vector $\gamma_{n,m}$ depend only of the two
``quantum'' numbers $n,m$ and of the cross section area $A$ of the
guide ($\gamma^{2}\sim 1/A$). The group velocity
$\partial\omega/\partial k_{x}$ of the wave in the guide is defined
by
$v_{g}=\left(c^{2}/\omega\right)\sqrt{\omega^{2}/c^{2}-\gamma^{2}}\simeq
c[1- \case{1}{2}c^{2}\gamma^{2}/\omega^{2}]$ and the travel time
$\Delta \tau $ by $L/v_{g}\simeq L[1+
\case{1}{2}c^{2}\gamma^{2}/\omega^{2}]/c$ which implies $\delta
t_{0}=\case{1}{2}L c\gamma^{2}/\omega^{2}$. In the moving case where
the cylinder possesses the velocity $v_{e}$ we can directly obtain
the condition given by Eq.~\ref{ultime3} (with $\gamma_{e}=1$) and
then we can deduce the group velocity of the sound in the guide with
the formula
\begin{eqnarray}
v=v_{e}+\frac{c-v_{e}}{1+\frac{c  \delta
t_{0}}{L}\left(1-\beta_{e}\right)} \label{fin1}.
\end{eqnarray}
This last equation give us the Fresnel result if we put the
effective sound index $n_{0}=1+c\delta t_{0}/L$. We can control the
self consistency of this calculation by observing that the
dispersion relation Eq.~\ref{dispersion} allows the definition of a
phase index $n_{\text{phase}}=c k/\omega\simeq
1-c^{2}\gamma_{n,m}^{2}/(2\omega^{2})$ which is equivalent to
Eq.~\ref{ultime3} when $\omega_{0}=0$ and $2\pi
N_{0}e^{2}/m=c^{2}\gamma_{n,m}^{2}/2$. This reveals a perfect
analogy between the sound wave propagating in a moving cylinder and
the light wave propagating in a
moving slab. It is then not surprising that the Fresnel result is correct in the two cases. \\
 The principal limitation of our deduction is contained in
the assumption expressed above for the slab example: $ \|\mathbf{
E}_{\text{after}}\|\simeq \|\mathbf{ E}_{\text{before}}\|$ i.~e. the
condition of no reflection supposing the perturbation on the motion
of the wave to be small. Nevertheless, the principal origin of the
Fresnel effect is justified in our scheme without the use of the
Einstein relativity principle.\\
We can naturally ask if the simple analogy proposed can not be
extended to a dense medium i.~e.~without the approximation of a weak
density $N_{0}$ or of a low reflectivity. In order to see that it is
indeed true we return to the electromagnetic theory and we suppose
an infinite moving medium like the one considered in the second
section. In the rest frame of the medium we can define a slab of
length $L_{0}$. The unique difference with the section 3 is that now
this slab is not bounded by two interfaces separating the atoms from
the vacuum but is surrounded by a continuous medium having the same
properties and moving at the same velocity $v_{e}$. In the
laboratory frame the length of the moving slab is
$L=L_{0}\gamma_{e}^{-1}$. We can write the time $\Delta \tau_{e}$
taken by a signal like a wave packet, a wave front or a plane of
constant phase to travel through the moving slab:
\begin{equation}
c\Delta\tau_{e}=n\cdot\left(L+v_{e}\Delta\tau_{e}\right).
\end{equation}The optical index $n$
can be the one defined in section 2 for the case of the Drude model
but the result is very general. We can now introduce a time $\delta
t_{0}$ such that Eq.~\ref{ultime3}, and consequently Eq.~\ref{fin},
are true \emph{by definition}. We conclude that this last equation
Eq.~\ref{fin} is equivalent to the relativistic Eq.~\ref{relativist}
if, and only if, we define the time $\delta t_{0}$ by the formula
\begin{equation}
\delta t_{0}=\left(n_{0}-1\right)\cdot L_{0}/c.\label{pursue}
\end{equation} In other terms we can always use
the analogy with a photon pursuing an atom since the general formula
Eq.~\ref{relativist} is true whatever the microscopic and
Electrodynamics model considered. In this model - based on a
retardation effect-  the absorbtion time $\delta t_{0}$ is always
given by Eq.~\ref{pursue}. \\
%corrections lignes +1, +4, +6, +7, +9, +10, +11, +12
This opens new perspectives when we
consider the problem of a sound wave propagating in an effective
moving medium. Indeed there are several situations where we can
develop a deep analogy between the propagation of sound and the
propagation of light. This implies that the conclusions obtained for
the Fresnel effect for light must to a large part be valid for
sound as well. This is in particular true if we consider an effective
meta material like the one
that is  going to be described now:\\
We consider a system of mirrors as represented in Fig.~2A,
at rest in the laboratory. A beam of light propagates
along the zigzag trajectory
$A_{0},B_{0},A_{1},...,A_{n},B_{n},...$. The length $A_{n}B_{n}$ is
given by $\sqrt{(L_{0}^{2}+D^{2})}$ where the distance $L_{0}$ and
$D$ are represented on the figure. The time $\Delta \tau_{0}$ spent
by a particle of light to move along $A_{n}B_{n}$ is then
$\sqrt{(L_{0}^{2}+D^{2})}/c$. We can equivalently define an
effective optical index $n_{0}$ such that we have
\begin{equation}
\frac{(L_{0}^{2}+D^{2})}{c^{2}}=\Delta
\tau_{0}^{2}=\frac{L_{0}^{2}n_{0}^{2}}{c^{2}}.
\end{equation}
This implies
\begin{equation}
n_{0}^{2}= 1 +\frac{D_{0}^{2}}{L_{0}^{2}}.
\end{equation}We consider now the same problem for a system of mirrors moving
with the velocity $v_{e}$. In order to be consistent with relativity
we introduce the reduced length $L=L_{0}\gamma_{e}^{-1}$. The beam
propagating along the path $A_{0},B_{0},A_{1},...,A_{n},B_{n},...$
must pursue the set of mirrors. We then define the travel time
$\Delta \tau_{e}$ along an elementary path $A_{n}B_{n}$ by
\begin{equation}
\frac{((L+v_{e}\Delta \tau_{e})^{2}+D^{2})}{c^{2}}=\Delta
\tau_{e}^{2}=\frac{((L+v_{e}\Delta \tau_{e})^{2}n^{2}}{c^{2}},
\end{equation} where $n$ is the effective optical index for the
moving medium. From this equation we deduce first $\Delta
\tau_{e}=(L/c)n/(1-v_{e}n/c)$ and then
\begin{equation}
n^{2}-1=(n_{0}^{2}-1)\frac{(1-\frac{v_{e}n}{c})^{2}}{(1-(\frac{v_{e}}{c})^{2})}
\end{equation} which finally give us the formula
\begin{equation}
\frac{c}{n}=\frac{c/n_{0}+v_{e}}{1+\frac{v_{e}}{cn_{0}}}
\end{equation}
We can again justify the Fresnel formula at the limit $v_{e}/c\ll1$.
\begin{figure}
\includegraphics[width=3in]{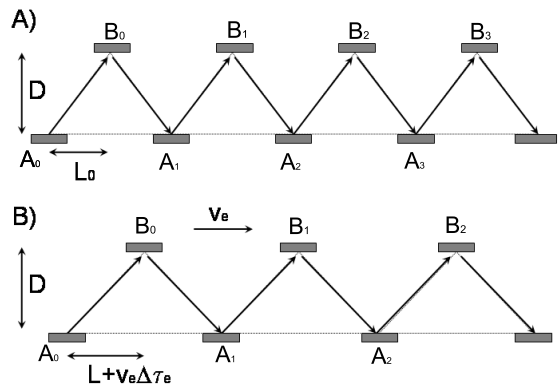}\\
\caption{An ideal meta material equivalent to a medium with an
effective index. A wave represented by an arrow propagates between
the mirrors $A_{0},B_{0},A_{1},...,A_{n},B_{n},...$. A) when the
mirrors are at rest in the laboratory the travel time
$\delta\tau_{0}=A_{n}B_{n}/c$ is dependent only on the distances
$L_{0}$ and $D$. B) When the mirrors move at the velocity $v_{e}$
relatively to the laboratory the travel time
$\delta\tau_{e}=A_{n}B_{n}/c$ is affected by the motion and depends
on $v_{e}$ as well as on $D$ and $L=L_{0}/\gamma_{e}$.}
\label{simple}
\end{figure}
 The simplicity of this model is such that it does not depend
on the physical properties of atoms, electrons and photons but only
on geometrical parameters. Clearly we can make the same reasoning
for a sound wave by putting $\gamma_{e}=1$. This still gives us the
Fresnel formula when we neglect terms equal or smaller than
$O\left(v_{e}^{2}/c^{2}\right)$. In addition this model allows us to
conclude that the essential element justifying the Fresnel-Fizeau
result is the emergence of a delay time -- a retardation effect--
when we consider the propagation of the signal at a microscopic or
internal level. The index $n$ which characterizes the macroscopic or
external approach is then just a way to define an effective velocity
without looking for a causal explanation of the retardation.\\
The essential message of our analysis is that by taking explicitly
into account the physical origin of the delay we can justify the
essence of the Fresnel-Fizeau effect in a non relativistic way. The
Fresnel-Fizeau effect is then a very general phenomenon. It is a
consequence of the conservation of energy and momentum and of the
constant value of the wave velocity
in vacuum or in the considered medium. The so called travel
condition (Eq.~\ref{condition}) which is a combination of these two
points can be compared to the usual demonstration for the Doppler
effect. In these two cases of light pulses pursuing a moving
particle the perturbation time $\delta t_{v_{e}} \simeq \delta
t_{0}\left(1-v_{e}/c\right)$ is a manifestation of the Doppler
phenomenon. It should be emphasized that the analogy between sound
and electromagnetic waves discussed in this article could be
compared to the similarities between sound wave and gravitational
waves discussed in particular by Unruh. On this subject and some
connected discussions concerning the acoustic Aharonov-Bohm effect
(that is related to the optical Aharonov-Bohm effect that follows
from the Fizeau effect) the reader should
consult \cite{Unruh1,Unruh2}.
 \section{Summary}
We have obtain the Fresnel-Fizeau formula using a perturbation
method based on the optical theorem and in a more general way by
%correction ligne +1
considering the physical origin of the refractive index. The
modification  of the speed of light in the medium appears then as a
result of a retardation effect due to the duration of the
interaction or absorbtion of light by the medium, and the
Fresnel-Fizeau effect, as a direct consequence of the medium's
flight in front of the light. These facts rely on the same origin as
%corrections lignes +1,2,3
the Doppler-Fizeau effect. We finally have shown that
 it is not correct to assume, as frequently done in the past, that a coherent and ``Newtonian
interpretation'' of these phenomena would be impossible.
 On the contrary, the results do not invalidate the derivation of the Fresnel-Fizeau effect based on the principle of relativity but clarify it. We observe indeed than all reasoning
is in perfect agreement with the principle of relativity. We must
emphasize  that even if the Fizeau/Fresnel effect is conceptually
divorced from relativity it strongly motivated Einstein's work (more
even than the Michelson and Morley result). The fact that the Fizeau
as well as the Michelson-Morley experiment can be justified so
%corrections +1,2,3,4,5
easily with special relativity clearly show the advantages of
Einstein's principle to obtain quickly the correct results.
Nevertheless, if we look from a dynamical point of view, as it is the
case here, this principle plays a role only for effects of order
$\mathbf{v}_{e}^{2}/c^{2}$ which however are not necessary to
justify the Fresnel formula.
\begin{acknowledgments}
The author acknowledges S.~Huant, M.~Arndt, J.~Krenn, D.~Jankowska
as well as the two anonymous referees for interesting and fruitful
discussions during the redaction process.
\end{acknowledgments}

%%%
%%
%%%
\appendix
 \section{}
%correction +1,8,9
 Maxwell's equations impose the continuity of the electric field on each interface of the slab.
 More precisely these boundary relations impose:
 $\mathbf{ E}_{\textrm{medium A}}\vert_{S}=\mathbf{ E}_{\textrm{medium B}}\vert_{S}$ where $S$ is one of
  the two moving interfaces separating vacuum and matter.
 Hence we obtain an equality condition between the two phases $\phi_{\textrm{medium A}}$ and $\phi_{\textrm{medium B}}$
 valid for all times at the interface. Let $\Phi_{1}=-i\omega\left(t-x/c\right)$ be the phase of the plane wave before the slab.
 In a similar way let $\Phi_{2}=-i\omega_{2}\left(t-n_{\omega_{2},\beta_{e}}x/c-\delta_{2}\right)$
 and $\Phi_{3}=-i\omega_{3}\left(t-x/c-\delta_{3}\right)$ be the phases in the slab and in vacuum after traversing the
 slab, respectively. In these expressions there appear two retardation constants, $\delta_{2,3}$ and the optical index of the slab.
On the first interface denoted by (I-II) we have $x=c\beta_{e}t$
and consequently
\begin{eqnarray}
\omega\left(1-\beta_{e}\right)t=
\omega_{2}\left(1-n\left(\omega,\beta_{e}\right)\beta_{e}\right)\nonumber\\
\cdot\left(t-\frac{\delta_{2}}{1-n\left(\omega_{2},\beta_{e}\right)\beta_{e}}\right),
\end{eqnarray}
 which is valid for each time and possesses the unique solution:
\begin{eqnarray}
\omega_{2}=\omega\frac{1-\beta_{e}}{1-n\left(\omega,\beta_{e}\right)\beta_{e}}&
, \delta_{2}= 0.
\end{eqnarray}
 Considering the second interface
(II-III) in a similar way we obtain the following conditions
\begin{eqnarray}
\omega_{3}=\omega_{2}\frac{1-n\left(\omega,\beta_{e}\right)\beta_{e}}{1-\beta_{e}}=\omega
\nonumber \\ \delta_{3}=
L\frac{n\left(\omega,\beta_{e}\right)-1}{c-n\left(\omega,\beta_{e}\right)c\beta_{e}}
\end{eqnarray} where the $2^{nd}$ equality is Eq.~\ref{ultime2}.
%%%
%%%%
\section{}
\begin{figure}[h]
  \includegraphics[width=3.5583in]{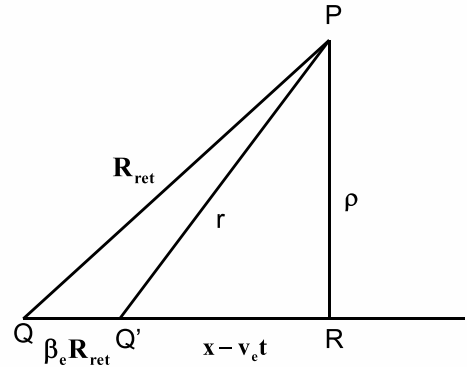}\\
  \caption{In this figure Q is the position of the particle at the
retarded time $t_{ret}$ and P is the observation point at the time
$t$. The particle moves uniformly on the the $x$ line QR following
the trajectory $v_{e}t$ and Q' is the position of the particle at
%corrections +1,2
the time $t$ separated of P by the distance $r$. In addition, if we
call R the projection of P on QR, then $x$ and $\rho$ are the
coordinates of the observation point in the plane of the figure.}
\label{retard}
 \end{figure}

Using geometrical considerations (see Fig.~\ref{retard}) we can
deduce the relation
\begin{equation}
R\left(t_{ret}\right)^{2}=\rho^{2}+\left(\beta_{e}R_{ret}+x-v_{e}t\right)^{2},
\end{equation}
which is equivalent after manipulations to the other:
\begin{equation}
\rho^{2}+\gamma_{e}^{2}\left(x-v_{e}t\right)^{2}=\left(1-\beta_{e}^{2}\right)\left(R_{ret}-\beta_{e}\gamma_{e}^{2}\left(x-v_{e}t\right)\right)^{2}.
\end{equation}
We can in a second step rewrite this equality as follows:
\begin{equation}
R_{ret}= \gamma_{e}^{2}\beta_{e}\left(x-v_{e}t\right)+
\gamma_{e}\sqrt{\rho^{2}+\gamma^{2}_{e}\left(x-v_{e}t\right)^{2}}
\end{equation}which is Eq.~\ref{cactus}.

\end{document}